\documentclass[twocolumn]{autart} 
\usepackage{amsfonts}
\usepackage{amssymb}
\usepackage{amsmath}                                                        
\usepackage{subfig}
\usepackage{epsfig}
\usepackage{graphicx}
\newtheorem{theorem}{Theorem}
\newcommand{\startproof}{\vspace{0.01cm} \noindent {\bf \em Proof: }}   
\newcommand{\finishproof}{\hfill $\blacksquare$}    
\newtheorem{assumption}{Assumption}     
\newtheorem{observation}{Observation}                                                

\newcommand{\eqdef}{\stackrel{\triangle}{=}}
\begin{document}
\begin{frontmatter}
\title{Robust Dynamic State Feedback Guaranteed Cost Control of Nonlinear Systems using Copies of Plant Nonlinearities}

\author[SEIT]{Obaid Ur Rehman}\ead{s.obaid.rehman@gmail.com},    
\author[SEIT]{Ian R. Petersen}\ead{i.r.petersen@gmail.com}, 
\address[SEIT]{School of Engineering and Information Technology, University of New South Wales, Canberra, Australia.} 
\thanks{This research was supported by the Australian Research Council.}
\begin{abstract}
This paper presents a systematic approach to the design of a robust dynamic state feedback controller using copies of the plant nonlinearities, which is based on the use of IQCs and minimax LQR control. The approach combines a linear state feedback guaranteed cost controller and copies of the plant nonlinearities to form a robust nonlinear controller.
\end{abstract}
\end{frontmatter}
\section{INTRODUCTION}
This paper presents a new approach to the constructive design of a robust nonlinear dynamic state feedback  controller using an integral quadratic constraint (IQC) approach. The idea of using a copy of the plant nonlinearity in the controller is used previously in the literature \cite{outputfeedback_petersen,GLover_Nonlinear,Chu_Nonlinear}. However in this paper, we apply a new methodology to  construct a controller which uses linear state feedback guaranteed cost control and copies of the plant nonlinearities to form a dynamic state feedback robust nonlinear controller. This approach provides robust performance in the case where uncertainties and nonlinearities are present in the plant. 
\section{System Definition}\label{sec:system}
Consider a class of uncertain nonlinear systems described by the following state equations:
\begin{small}
\begin{equation}
\label{eqsystem}
\begin{split}
\dot{x}(t)&=A x(t)+ \sum_{j=1}^k \check{B}_{1,j} \xi_{1,j}(t)+ \sum_{i=1}^g \bar{B}_{1,i}\mu_i (t)\\
&\qquad \qquad+B_2(t) u(t);\quad x(0)=x_0;\\
\zeta_{1,j}(t)&=\check{C}_{1,j} x(t)+\check{D}_{1,j}u(t);~j=1,\cdots,k,\\
\nu_i (t)&=\bar{C}_{1,i} x(t)+\bar{D}_{1,i}u(t);~i=1,\cdots,g,
\end{split}
\end{equation}
\end{small}

\noindent where $x(t)\in \Re^{{n}}$ is the state, $u(t)\in \Re^{{m}}$ is the control input, $\zeta_{1,1}(t)\in \Re^{q_1}$, $\zeta_{1,2}(t)\in \Re^{q_2}$, $\cdots$, $\zeta_{1,k}(t)\in \Re^{q_k}$ are the uncertainty outputs, $\xi_{1,1}(t)\in \Re^{p_1}$, $\xi_{1,2}(t)\in \Re^{p_2}$, $\cdots$, $\xi_{1,k}(t)\in \Re^{p_k}$ are the uncertainty inputs, $\nu_1(t)\in \Re^{h_1}, \cdots, \nu_g(t)\in \Re$ are the nonlinearity outputs, $\mu_1(t)\in \Re, \cdots, \mu_g(t)\in \Re$ are the nonlinearity inputs.
Also,
\begin{equation}
\begin{split}
\xi_{1,j}(t)&=\phi_j(\zeta_{1,j}(t),t);\quad \forall j=1,\cdots,k.\\
\end{split}
\end{equation}
The nonlinearity inputs and outputs are related as follows:
\begin{equation}
\label{eqnon1}
\mu_i(t)=\psi_i (\nu_i(t),t) \quad \forall i=1,2, \cdots, g,
\end{equation}
where the nonlinear functions $\psi_i(\cdot)$ satisfy the following generalized monotonicity conditions:
\begin{equation}
\label{eqmon1}
[\psi_i(\nu_1)-\psi_i(\nu_2)\quad \nu_1-\nu_2]N_i\left[
\begin{array}{c}
\psi_i(\nu_1)-\psi_i(\nu_2)\\
\nu_1-\nu_2
\end{array}\right]\geq0
\end{equation}
for all $\nu_1 \in \Re $, $\nu_2 \in \Re$ and $i=1,2, \cdots, g$. Also, $N_i\in \Re^{2\times 2}$ are given symmetric matrices representing the monotonicity or global Lipschitz conditions; see \cite{outputfeedback_petersen}. 
Furthermore, we assume that
\begin{equation}
\label{eqmon2}
\psi_i(0)=0\quad \forall~ i=1,2,\cdots,g
\end{equation}
and the uncertainty in the system satisfies the following integral quadratic constraints, (see \cite{IP}):
\begin{small}
\begin{equation}
\label{eqIQC1}
\int_0^{T}  [\xi_{1,j}(t)' \quad \zeta_{1,j}(t)'] M_{1,j} \left[
\begin{array}{c}
\xi_{1,j}(t)\\
\zeta_{1,j}(t)
\end{array}\right] dt +x(0)^T S_{1,j} x(0)\geq 0,
\end{equation}
\end{small}
\begin{figure}[t]
\begin{center}
\epsfig{file=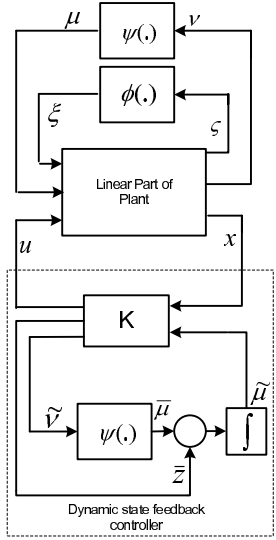, scale=0.6}
\caption{Nonlinear system with a nonlinear dynamic state feedback  controller.}
\label{fig:nonlinear}
\end{center}
\end{figure}
\noindent for all $j=1,2, \cdots, k$. Here the $M_{1,j}$ are given symmetric matrices and the $S_{1,j}$ are given positive definite matrices. Let us define
\begin{small}
\begin{equation}
\label{eqnx}
\tilde{x}(t)\eqdef
\left[
\begin{array}{c}
x(t) \\
\tilde{\mu}_1(t)\\
\vdots\\
\tilde{\mu}_g(t)\\
\end{array}\right];\quad
\tilde{u}(t)\eqdef
\left[
\begin{array}{c}
u(t) \\
\tilde{\nu}_1(t)\\
\vdots\\
\tilde{\nu}_g(t)\\
\bar{z_1}\\
\vdots\\
\bar{z_g}
\end{array}\right].
\end{equation}
\end{small}

\noindent Hence the class of controllers considered here are nonlinear dynamic state feedback controllers which contain a copy of the plant nonlinearities (see Fig. \ref{fig:nonlinear});
\begin{small}
\begin{equation}
\label{eqnestimator}
\tilde{u}= K \tilde{x},
\end{equation}
where
\begin{equation}
\label{eqdotmu}
\dot{\tilde{\mu}}_i =\bar{\mu}_i+\bar{z}_i,
\end{equation}
and
\begin{equation}
\label{eqnon2}
\begin{split}
\bar{\mu}_i(t)&=\psi_i (\tilde{\nu}_i(t)), \quad i=1,2, \cdots, g.
\end{split}
\end{equation}
\end{small}

\noindent Also $K$ is a controller gain matrix. 
The following IQCs, which follow from (\ref{eqmon1}), are satisfied:
\begin{align}
\label{eqIQC2}
E&\big\{\int_0^T[\mu_i-\tilde{\mu}_i \quad \nu_i-\tilde{\nu}_i] N_i\left[
\begin{array}{c}
{\mu}_i-\tilde{\mu}_i\\
\nu_i-\tilde{\nu}_i
\end{array}\right]dt\nonumber \\ 
&+x(0)'\breve{S}_{i,1}x(0)\big\}\geq0;\\
\label{eqIQC3}
E&\big\{\int_0^T[\mu_i \quad \nu_i] N_i\left[
\begin{array}{c}
{\mu}_i\\
\nu_i
\end{array}\right]dt+x(0)'\breve{S}_{i,2}x(0)\big\}\geq0;\\
\label{eqIQC4}
E&\big\{\int_0^T[\tilde{\mu}_i \quad \tilde{\nu}_i] N_i\left[
\begin{array}{c}
\tilde{\mu}_i\\
\tilde{\nu}_i
\end{array}\right]dt+x(0)'\breve{S}_{i,3}x(0)\big\}\geq0;
\end{align}
for all $i=1, \cdots , g$. Here the $\breve{S}_{i,1}$, $\breve{S}_{i,2}$, $\breve{S}_{i,3}$ are any positive definite matrices.  
Now, we first move the controller nonlinearities (\ref{eqnon2}) into the plant description and introduce new notation as follows:
\begin{small}
\begin{equation}
\label{eqnstate}
\begin{split}
&\tilde{B}_{2,i} \eqdef \left[
\begin{array}{cc}
\bar{B}_{1,i} & 0 \\
0 & I
\end{array}\right];\quad \tilde{B}_{2} \eqdef \left[
\begin{array}{ccc}
\bar{B}_2 & 0 & 0\\
0 & 0 & I
\end{array}\right];\\
&\tilde{\xi}_{2,i}=\left[
\begin{array}{c}
\mu_i \\
\tilde{\mu}_i
\end{array}\right];\quad \tilde{\zeta}_{2,i}=\left[
\begin{array}{c}
\nu_i \\
\tilde{\nu}_i
\end{array}\right];\quad \tilde{C}_{1,j}=\left[
\begin{array}{ccc}
\check{C}_{1,j} & 0
\end{array}\right];\\
&\tilde{B}_{1,j} \eqdef \left[
\begin{array}{c}
\check{B}_{1,j} \\
0
\end{array}\right];\quad \tilde{C}_{2,i}=\left[
\begin{array}{ccc}
\bar{C}_{1,i} & 0 \\
0 & 0 
\end{array}\right];\\
&\tilde{D}_{2,i}=\left[
\begin{array}{ccc}
\bar{D}_{1,i} & 0 \\
0 & 0 
\end{array}\right];\quad \tilde{D}_{1,j}=\left[
\begin{array}{c}
\check{D}_{1,j} \\
0
\end{array}\right],
\tilde{A}=\left[
\begin{array}{cc}
A & 0 \\
0 & 0
\end{array}\right],
\end{split}
\end{equation}
\end{small}

\noindent for all $i=1,\cdots,g$ and $j=1,\cdots,k$. Using the above notation and (\ref{eqnx}), a new system can be written as follows:
\begin{equation}
\label{eqsystemc1}
\begin{split}
\dot{\tilde{x}}&=\tilde{A}\tilde{x}+\sum_{j=1}^k \tilde{B}_{1,j} \xi_{1,j} +\sum_{i=1}^g \tilde{B}_{2,i} \tilde{\xi}_{2,i} + \tilde{B}_2\tilde{u}\\
\zeta_{1,j}&=\tilde{C}_{1,j}\tilde{x}+\tilde{D}_{1,j} \tilde{u};~\tilde{\zeta}_{2,i}=\tilde{C}_{2,i}\tilde{x}+\tilde{D}_{2,i} \tilde{u},
\end{split}
\end{equation}
$\forall~ i=1,\cdots,g$ and $j=1,\cdots,k$. Also, the IQCs (\ref{eqIQC2})--(\ref{eqIQC4}) for the nonlinear uncertainty terms can be written as follows:
\begin{equation}
\label{eqIQC}
\int_0^{T}  [\tilde{\xi}_{2,i}' \quad \tilde{\zeta}_{2,i}'] \tilde{M}_{i,p} \left[
\begin{array}{c}
\tilde{\xi}_{2,i}\\
\tilde{\zeta}_{2,i}
\end{array}\right] dt +x(0)^T \tilde{S}_{i,p} x(0)\geq 0,
\end{equation}
for $i=1,2, \cdots, g,$ $p=1,2,3$, where $\tilde{M}_{i,p}$ and $\breve{S}_{i,p}$ are positive definite matrices. We consider the following cost functional associated with the system (\ref{eqsystemc1}):
\begin{equation}
\label{eqnfcost}
\begin{split}
\tilde{J}(\cdot)&=\int_0^\infty(\tilde{x}^T \tilde{R} \tilde{x}+\tilde{u}^T \tilde{G} \tilde{u})dt,
\end{split}
\end{equation}
where $\tilde{R}\in \Re^{n \times n}$ and $\tilde{G}\in \Re^{m \times m}$ are positive-definite symmetric matrices. 
\begin{observation}\label{Obs1}
It is observed that the nonlinearities (\ref{eqnon1}) and (\ref{eqnon2}) satisfy the IQCs (\ref{eqIQC2})--(\ref{eqIQC4}). Hence, it follows that if the linear uncertain system (\ref{eqsystemc1}) and (\ref{eqIQC}), with the linear controller (\ref{eqnestimator}), leads to an upper bound on the cost function (\ref{eqnfcost}) then the same controller (\ref{eqnestimator}) will yield the same upper bound for the uncertain system (\ref{eqsystem}), (\ref{eqnon1}), (\ref{eqIQC1}) and (\ref{eqnon2}). Furthermore, it follows from the above discussion that the system (\ref{eqsystem}), (\ref{eqnon1}), (\ref{eqIQC1}) with controller (\ref{eqnestimator}) and (\ref{eqnon2}) will also lead to the same upper bound on the cost.
\end{observation}
\section{Main Results}\label{sec:main}
We first write the IQC (\ref{eqIQC}) in the following form which is parametrized by a set of multipliers $\lambda_i>0$ for $i=1,2,\cdots,g$:
\begin{equation}
\label{eqIQCM}
\int_0^{t_i}  [\tilde{\xi}_{2,i}' \quad \tilde{\zeta}_{2,i}'] \tilde{M}_{i}(\lambda_i) \left[
\begin{array}{c}
\tilde{\xi}_{2,i}\\
\tilde{\zeta}_{2,i}
\end{array}\right] dt +x(0)^T \tilde{S}_{i,p}(\lambda_i) x(0)\geq 0,
\end{equation}
where $\tilde{M}_{i}(\lambda_i)=\sum_{p=1}^3 \lambda_{i,p} \tilde{M}_{i,p}$, $\tilde{S}_{i}(\lambda_i)=\sum_{p=1}^3 \lambda_{i,p} \tilde{S}_{i,p}$. Furthermore, $\lambda_i=[\lambda_{i,1},~\lambda_{i,2}, \lambda_{i,3}]$ where all $\lambda_i\in \Gamma=\{\lambda_i \in \Re^3 : \lambda_{i,p}\geq 0 \forall i, p \}$. Note that we only consider those $\lambda_i \in \Gamma$ which satisfy the following condition
\begin{equation}
\label{eqcondq}
\Pi(\tilde{M}_i(\lambda_i))=2; ~det U_{i,11}(\tilde{M}_i(\lambda_i))\neq0
\end{equation}
where  $\Pi(\cdot)$ represents the number of negative eigenvalues of $\tilde{M}(\lambda_i)$  and $U_{11}(\tilde{M}_i(\lambda_i))$ is the matrix of eigenvectors corresponding to the negative eigenvalues. 
If condition (\ref{eqcondq}) is satisfied, then there exists a matrix $T_i$ such that the matrix $T_i'(\tilde{M}_i(\lambda))T_i$ is a diagonal matrix 
\begin{equation}
T_i'(\tilde{M}_i(\lambda))T_i=\left[
\begin{array}{cc}
-\mathbf{I}_{2\times 2} & 0 \\
0 & \mathbf{I}
\end{array}\right].
\end{equation}
We define $T_i=U(\tilde{M}_i(\lambda_i)) D_i U(\tilde{M}_i(\lambda_i))^{-1}$
where $D_i(\tilde{M}_i(\lambda_i))=\left[
\begin{array}{cc}
D_{1,i} & 0 \\
0 & D_{2,i}
\end{array}\right]$, $D_{1,i}=diag[\sqrt{-\bar{\sigma}_{1,i}}, \sqrt{-\bar{\sigma}_{2,i}}]$ and  $D_{2,i}= diag [\sqrt{{\sigma}_{1,i}},\sqrt{{\sigma}_{2,i}}]$. Here ${\bar{\sigma}_{1,i}},{\bar{\sigma}_{2,i}}$ are the negative eigenvalues and ${{\sigma}_{1,i}},{{\sigma}_{{2,i}}}$ are the positive eigenvalues of the matrix $\tilde{M}_i(\lambda)$.
Now a change in variables is introduced as follows:
\begin{equation}
\label{eqchvar1}
\left[
\begin{array}{c}
\tilde{\xi}_{2,i}(t)\\
\tilde{\zeta}_{2,i}(t)
\end{array}\right]=T_i \left[
\begin{array}{c}
\bar{\xi}_{2,i}(t)\\
\bar{\zeta}_{2,i}(t)
\end{array}\right];
\end{equation}
\begin{equation}
\label{eqchvar2}
\left[
\begin{array}{c}
\bar{\xi}_{2,i}(t)\\
\bar{\zeta}_{2,i}(t)
\end{array}\right]=T_i^{-1}\left[
\begin{array}{c}
\tilde{\xi}_{2,i}(t)\\
\tilde{\zeta}_{2,i}(t)
\end{array}\right]=\left[
\begin{array}{cc}
\tilde{T}_{11} & \tilde{T}_{12}\\
\tilde{T}_{21} & \tilde{T}_{22}
\end{array}\right] \left[
\begin{array}{c}
\tilde{\xi}_{2,i}(t)\\
\tilde{\zeta}_{2,i}(t)
\end{array}\right].
\end{equation}
The IQCs (\ref{eqIQCM}) for a given $i\in {1,\cdots,g}$ can now  be modified by incorporating new variables as given below (from now on we remove the argument ($t$) from the equations wherever possible for the sake of brevity):
\begin{small}
\begin{equation}
\label{eqnIQC}
\begin{split}
\int_0^T[\bar{\xi}_{2,i}' \quad \bar{\zeta}_{2,i}']\left[
\begin{array}{cc}
-\textbf{I} & 0\\
0 & \textbf{I}
\end{array}\right]\left[
\begin{array}{c}
\bar{\xi}_{2,i}\\
\bar{\zeta}_{2,i}
\end{array}\right]dt
+x(0)'\tilde{S}_{2,i}(\lambda)x(0)\geq0.
\end{split}
\end{equation}\end{small}
Hence,
\begin{small}
\begin{equation}
\label{eqnnIQC}
\int_0^T\Vert\bar{\xi}_{2,i}'\Vert^2- \Vert \bar{\zeta}_{2,i}'\Vert^2 dt -x(0)'\bar{S}_{2,i}x(0)\leq0.
\end{equation}\end{small}  

We have $\bar{\xi}_{2,i}=\tilde{T}_{11}\tilde{\xi}_{2,i}+\tilde{T}_{12} \tilde{\zeta}_{2,i}$ and $\breve{\zeta}_{2,i}=\tilde{T}_{21}\tilde{\xi}_{2,i}+\tilde{T}_{22}\tilde{\zeta}_{2,i}$, which imply the following relation:
\begin{equation}
\label{eqnxizeta}
\begin{split}
\tilde{\xi}_{2,i}&=\tilde{T}_{11}^{-1}(\bar{\xi}_{2,i}-\tilde{T}_{12}\tilde{\zeta}_{2,i});\\  \breve{\zeta}_{2,i}&=\tilde{T}_{21}\tilde{T}_{11}^{-1}(\bar{\xi}_{2,i}-\tilde{T}_{12}\tilde{\zeta}_{2,i})+\tilde{T}_{22}\tilde{\zeta}_{2,i}.
\end{split}
\end{equation}
Hence, we obtain
\begin{equation}
\left[
\begin{array}{c}
\tilde{\xi}_{2,i}(t)\\
\bar{\zeta}_{2,i}(t)
\end{array}\right]=\left[
\begin{array}{cc}
\tilde{T}_{11}^{-1} & -\tilde{T}_{11}^{-1}\tilde{T}_{12}\\
\tilde{T}_{21}\tilde{T}_{11}^{-1} & \tilde{T}_{22}-\tilde{T}_{21}\tilde{T}_{11}^{-1}\tilde{T}_{12}
\end{array}\right] \left[
\begin{array}{c}
\bar{\xi}_{2,i}(t)\\
\tilde{\zeta}_{2,i}(t)
\end{array}\right].
\end{equation}
Substituting for $\tilde{\xi}$ and $\bar{\zeta}$ into (\ref{eqsystemc1}) gives the following dynamical system:
\begin{equation}
\label{eqC1system}
\begin{split}
\dot{\tilde{x}}&=\bar{A} \tilde{x}+ \sum_{j=1}^k \tilde{B}_{1,j} {\xi}_{1,j}+ \sum_{i=1}^g \bar{B}_{2,i} \bar{\xi}_{2,i}+\bar{B}_{2}\tilde{u} ;\\
\tilde{\zeta}_{1,j}&=\tilde{C}_{1,j}  \tilde{x}+\tilde{D}_{1,j} \tilde{u},\bar{\zeta}_{2,i}=\bar{C}_{2,i}  \tilde{x}+\bar{D}_{2,i} \tilde{u}+\bar{D}_{11,i} \bar{\xi}_{2,i};
\end{split}
\end{equation}
for all $j=1,\cdots, k$, and $i=1,\cdots,g$ where
\begin{small}
\begin{equation}
\label{eqdefAbar}
\begin{split}
\bar{A}&\eqdef \tilde{A}- \sum_{i=1}^g \tilde{B}_{2,i} \tilde{T}_{11}^{-1} \tilde{T}_{12} \tilde{C}_{2,i},~\bar{B}_{2,i}\eqdef \sum_{i=1}^g \tilde{B}_{2,i} \tilde{T}_{11}^{-1},\\
\bar{B}_2 &\eqdef -\sum_{i=1}^g \tilde{B}_{2,i} \tilde{T}_{11}^{-1} \tilde{T}_{12} \tilde{D_1}+\tilde{B}_2,~\bar{D}_{11,i}\eqdef \tilde{T}_{21} \tilde{T}_{11}^{-1},\\
\bar{C}_{2,i}& \eqdef (\tilde{T}_{22}-\tilde{T}_{21}\tilde{T}_{11}^{-1}\tilde{T}_{12})\tilde{C}_{2,i},\bar{D}_{2,i}\eqdef (\tilde{T}_{22}-\tilde{T}_{21}\tilde{T}_{11}^{-1}\tilde{T}_{12})\tilde{D}_{2,i}.
\end{split}
\end{equation}
\end{small}

\noindent Also in order to deal with the $D_{11,i}$ terms in (\ref{eqC1system}) we use  standard loop shifting ideas \cite{outputfeedback_petersen,Basar_book} where we require that the following condition is satisfied for all $j=1,\cdots, k$ and $i=1,\cdots,g$: 
\begin{equation}
\label{eqcondD1}
\bar{D}_{11,i}' \bar{D}_{11,i}<I.
\end{equation}
For this purpose, we first define the following quantities:
\begin{equation}
\label{eqphis}
\begin{split}
\Phi_i&=I-\bar{D}_{11,i}' \bar{D}_{11,i}>0,~\bar{\Phi}_i=I-\bar{D}_{11,i} \bar{D}_{11,i}'>0.
\end{split}
\end{equation}
By using the definition in (\ref{eqphis}), we define the transformed uncertainty inputs and outputs as follows:
\begin{equation}
\label{eqndef}
\begin{split}
\check{\xi}_{2,i} &\eqdef \Phi_i^{1/2}\bar{\xi}_{2,i}-\Phi_i^{-1/2} \bar{D}_{11,i}'[\bar{C}_{2,i} x +\bar{D}_{2,i}\tilde{u}];\\
\check{\zeta}_{2,i} &\eqdef \bar{\Phi}_i^{-1/2} [\bar{C}_{2,i} x +\bar{D}_{2,i}\tilde{u}].
\end{split}
\end{equation}
Hence, $\bar{\xi}_{2,i}=\Phi_i^{-1/2}\check{\xi}_{2,i}+\Phi_i^{-1} \bar{D}_{11,i}'[\bar{C}_{2,i} x +\bar{D}_{2,i}\tilde{u}]$
and (\ref{eqnnIQC}) can be rewritten as follows for all $i=1,\cdots,g$:
\begin{small}
\begin{equation}
\label{eqfIQC}
\int_0^T\Vert\check{\xi}_{2,i}'\Vert^2- \Vert \check{\zeta}_{2,i}'\Vert^2 dt -x(0)'\tilde{S}_{2,i}x(0)\leq0.
\end{equation}
\end{small} 
Now, we re-write (\ref{eqC1system}) using (\ref{eqndef}) as follows:
\begin{equation}
\label{eqC2system}
\begin{split}
\dot{\tilde{x}}&=\check{A} \tilde{x}+ \sum_{j=1}^k \tilde{B}_{1,j} {\xi}_{1,j}+ \sum_{i=1}^g \bar{B}_{2,i} \check{\xi}_{2,i}+\check{B}_{2}\tilde{u} ;\\
{\zeta}_{1,j}&=\tilde{C}_{1,j}  \tilde{x}+\tilde{D}_{1,j} \tilde{u},~\check{\zeta}_{2,i}=\check{C}_{2,i}  \tilde{x}+\check{D}_{2,i} \tilde{u};\\
\check{\zeta}_{2,i}&=\bar{\Phi}_i^{-1/2}\bar{C}_{2,i}\tilde{x}+\bar{\Phi}_i^{-1/2}\bar{D}_{2,i}\tilde{u};
\end{split}
\end{equation}
for all $j=1,\cdots, k$, and $i=1,\cdots,g$ where
\begin{small}
\begin{equation}
\label{eqdefAbreve}
\begin{split}
\check{A}&\eqdef \bar{A}+\sum_{i=1}^g \bar{B}_{2,i}\bar{D}_{11,i}'\Phi_i^{-1} \bar{C}_{2,i},~\check{B}_{2,i}\eqdef \bar{B}_{2,i}\Phi_i^{-1/2};\\
\check{B}_{2}&\eqdef \sum_{i=1}^g \bar{B}_{2,i}\bar{D}_{11,i}'\bar{\Phi}_i^{-1} \bar{D}_{2,i}+\bar{B}_{2};\\
\check{C}_{2,i}& \eqdef \bar{\Phi}_i^{-1/2}\bar{C}_{2,i},~\check{D}_{2,i} \eqdef \bar{\Phi}_i^{-1/2}\bar{D}_{2,i}.\\
\end{split}
\end{equation}
\end{small}

\noindent In order to obtain a bound on the cost function (\ref{eqnfcost}), we design a guaranteed cost controller for the system (\ref{eqC2system}). The theory of guaranteed cost controllers can be found in \cite{IP}. 
In order to apply a guaranteed cost controller of the form (\ref{eqnestimator}), a parameter dependent algebraic Riccati equation is required to be solved for different values of the multipliers $\tau_{1,j}>0$ and $\lambda_i \in \Gamma$ for all $j=1,\cdots,k$  and $i=1,\cdots,g$. This Riccati equation is given below:
\begin{small}
\begin{equation}
\label{eqARE}
\begin{split}
&(\check{A}-\check{B}_2 G_{\tau}^{-1} D_{\tau} C_{\tau}) X_\tau+ X_\tau (\check{A}-\check{B}_2 G_{\tau}^{-1} D_{\tau} C_{\tau})^T\\
&-X_\tau \tilde{B}_2 G_{\tau}^{-1} \tilde{B}_2^T X_\tau + X_\tau \tilde{B}_{2\tau} \tilde{B}_{2\tau}^T X_\tau\\
&+ C_\tau^T (I-D_\tau G_{\tau}^{-1} D_\tau^T) C_\tau=0;
\end{split}
\end{equation}
\end{small}
where
\begin{small}
\begin{equation}
\label{eqdefAREs}
\begin{split}
&C_\tau=\left[
\begin{array}{c}
R^{1/2}\\
0\\
\sqrt{\tau_{1,1}} \tilde{C}_{1,1}\\
\vdots\\
\sqrt{\tau_{1,k}} \tilde{C}_{1,k}\\
 \check{C}_{2,1}\\
\vdots\\
 \check{C}_{2,g}\\
\end{array}\right],D_\tau=\left[
\begin{array}{c}
0\\
G^{1/2}\\
\sqrt{\tau_{1,1}} \tilde{D}_{1,1}\\
\vdots\\
\sqrt{\tau_{1,k}} \tilde{D}_{1,k}\\
 \check{D}_{2,1}\\
\vdots\\
 \check{D}_{2,g}\\
\end{array}\right],G_\tau=D_\tau^T D_\tau,\\
&B_{2\tau} = \left[
\begin{array}{cccccc}
\frac{\tilde{B}_{1,1}}{\sqrt{\tau_{1,1}}} &  \cdots & \frac{\tilde{B}_{1,k}}{\sqrt{\tau_{1,k}}} & \check{B}_{2,1} &  \cdots & \check{B}_{2,g}
\end{array}\right].
\end{split}
\end{equation}
\end{small}

\noindent The parameters $\tau_{1,1},\cdots,\tau_{1,k}$ and $\lambda_i \in \Gamma$ are chosen such that the Riccati equation (\ref{eqARE}) has a positive definite solution $X_\tau >0$.
\begin{assumption}\label{Ass1}
For any $\tau_{1,j}>0$ and $\lambda_i \in \Gamma$ $\forall~j=1,\cdots,k$ and $i=1,\cdots,g$ satisfying conditions (\ref{eqcondq}), (\ref{eqcondD1}), the following conditions are satisfied:
\begin{enumerate}
\item The Riccati equation (\ref{eqARE}) has a symmetric nonnegative definite solution $X_\tau$.
\end{enumerate}
\end{assumption}
\vspace{0.2mm}
If the conditions (\ref{eqcondq}), (\ref{eqcondD1}) along the Assumption \ref{Ass1} are satisfied then a controller of the form (\ref{eqnestimator}) for the system (\ref{eqC2system}) can be obtained as follows:
\begin{equation}
\label{eqcontrol2}
\tilde{u}=-G_{\tau}^{-1}[\check{B}_2^T X_\tau + D_\tau^T C_\tau]\tilde{x}
\end{equation}
and hence we define $K\eqdef-G_{\tau}^{-1}[\check{B}_2^T X_\tau + D_\tau^T C_\tau]$.
The corresponding bound on the cost function is obtained as follows:
\begin{equation}
\label{eqbound}
J\leq \tilde{x}_0^T X_\tau \tilde{x}_0 + \sum_{j=1}^k \tau_{1,j} \tilde{x}_0^T S_{1,j} \tilde{x}_0 + \sum_{i=1}^g  \tilde{x}_0^T \tilde{S}_{2,i} \tilde{x}_0.
\end{equation}
\begin{theorem}
Suppose there exist constants $\tau_{1,j}>0$ and vectors $\lambda_i \in \Gamma$ $\forall~j=1,\cdots,k$ and $i=1,\cdots,g$ such that conditions (\ref{eqcondq}), (\ref{eqcondD1}), along with Assumption 1 are satisfied. Then:
\begin{enumerate}
\item If the controller defined by (\ref{eqARE}), (\ref{eqdefAREs}), and (\ref{eqcontrol2}) is applied to the uncertain system defined by (\ref{eqfIQC}), (\ref{eqC2system}), (\ref{eqfIQC}), then the cost functional (\ref{eqnfcost}) satisfies the bound $\tilde{J}(u(\cdot))\leq V_\tau$.
\item If the controller defined by (\ref{eqARE}), (\ref{eqdefAREs}), and (\ref{eqcontrol2}) is applied to the  uncertain system defined by (\ref{eqnnIQC}), (\ref{eqC1system}),  then the cost functional (\ref{eqnfcost}) satisfies the bound $\tilde{J}(u(\cdot))\leq V_\tau$.
\end{enumerate}
\end{theorem}
\vspace{0.9mm}
\startproof
The first part of the theorem follows directly from the main results of \cite{IP} [Theorem 5.3.1]. The second part of the theorem is a result of  condition (\ref{eqcondD1}) which allows for the system (\ref{eqnnIQC}), (\ref{eqC1system}) to be written in the form  (\ref{eqfIQC}), (\ref{eqC2system}) and hence application of the result in \cite{IP} [Theorem 5.3.1] to this system will result in $\tilde{J}(u(\cdot))\leq V_\tau$ as noted in Observation \ref{Obs1}.
 \finishproof
\begin{theorem}\label{th1}
Suppose there exist constants $\tau_{1,j}>0$ and vectors $\lambda_i \in \Gamma$ $\forall~j=1,\cdots,k$ and $i=1,\cdots,g$ such that conditions (\ref{eqcondq}), (\ref{eqcondD1}), along with Assumption 1 are satisfied.  If the nonlinear controller defined by (\ref{eqnestimator}), (\ref{eqnon2}), (\ref{eqnstate}) is applied to the nonlinear uncertain system defined by (\ref{eqsystem}), (\ref{eqnon1}), (\ref{eqIQC1}), then the cost functional (\ref{eqnfcost}) satisfies the bound $\tilde{J}(u(\cdot))\leq V_\tau$.
\end{theorem}
\vspace{0.25mm}
\startproof
The result directly follows from part (ii) of Theorem \ref{th1} and the construction of the IQC (\ref{eqnnIQC}), the system (\ref{eqC1system}) and the controller (\ref{eqnestimator}) along with the discussion in Observation \ref{Obs1}.
\finishproof
\section{Illustrative Example}\label{sec:example}
\begin{figure}[t]
\hfill
\begin{center}
\epsfig{file=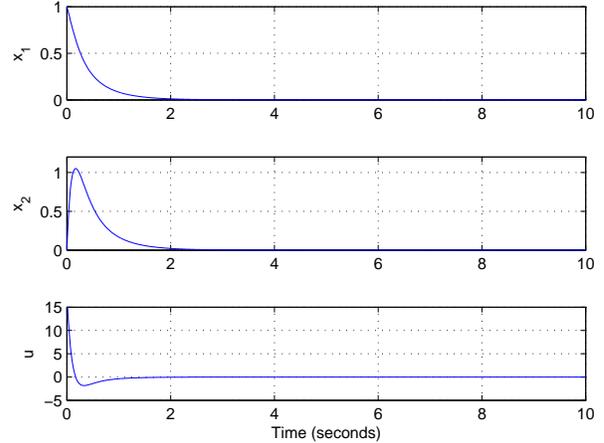, scale=0.6}
\caption{Compressor states $x_1$, $x_2$ and control input $u$.}
\label{fig:simulation}
\end{center}
\end{figure}

\vspace{0.25mm}
An example of state feedback control of axial compressor surge is considered in \cite{outputfeedback_petersen,Arcak_example,Krstic_Adaptive} and is given as follows:
\begin{small}
\begin{equation}
\dot{x}_1=-x_2-\frac{3}{2} x_1^2-\frac{1}{2}x_1^3, \quad \dot{x}_2=u;
\end{equation}
\end{small}

\noindent where $x_1$ and $x_2$ are the system states, and $u$ is the control input. In order to obtain a nonlinearity which is monotonic and sector bounded, we add a linear function to the nonlinearity. We also add an additional uncertainty satisfying an IQC for robustness purposes. Hence, we obtain
\begin{small} 
\begin{equation}
\begin{split}
\dot{x}_1&=(\frac{3}{2}x_1-x_2-\mu_1),~\dot{x}_2=(u+\xi_{1,1});\\
\zeta_{1,1}&=0.1 x_1; ~\nu_1=x_1;
\end{split}
\end{equation}
\end{small}

\noindent where $\mu_1=\psi(\nu_1)=\frac{3}{2}\nu_1+\frac{3}{2}\nu_1^2+\frac{1}{2}\nu_1^3$. The uncertainty input $\xi_{1,1}$ satisfies the IQC 
$\int_0^T \xi_{1,1}^2 dt \leq \int_0^T \zeta_{1,1}^2 dt$ for all $T>0$. We solve the algebraic Riccati equation (\ref{eqARE}) for the steady state stabilizing solution for possible values of $\tau_{1,1}>0$ and $\lambda_1\geq0$ satisfying conditions (\ref{eqmon1}), (\ref{eqmon2}), (\ref{eqcondD1}) along with Assumption 1 and considering the IQCs (\ref{eqIQC1}), (\ref{eqIQC2})-(\ref{eqIQC4}). These values of $\tau_{1,1}>0$ and $\lambda_1=[\lambda_{1,1},~\lambda_{1,2},~\lambda_{1,3}]$ are chosen so that steady state cost bound (\ref{eqbound}) is minimized.   The value of bound on the cost functional (\ref{eqnfcost}) for an initial condition of $x_1(0)=1$ and $x_2(0)=0$ is obtained as $V_\tau= 0.5772$ for the following values of the parameters:
\begin{small}
\[
\begin{split}
\lambda_{1,1}=&1 , ~\lambda_{1,2}=0.1,~\lambda_{1,3}=.12,~ \tau_{1,1}=0.15.
\end{split}
\]
\end{small}
The cost bound obtained using this scheme is lower than the cost bound obtained in \cite{outputfeedback_petersen} for the same example. This is expected as in the state feedback design we assume that all states are available for measurement. The nonlinear system with the nonlinear state feedback controller has also been simulated using  the above initial conditions and by assuming $\xi_{1,1}=0$. The result of the simulation is presented in Fig. \ref{fig:simulation}. It is observed that the control system performance is satisfactory.
\vspace{1mm}
\bibliographystyle{IEEEtran}

\end{document}